\documentclass[prl,twocolumn,letterpaper,superscriptaddress,reprint]{revtex4-1}

\usepackage{graphicx}
\usepackage{pslatex}
\usepackage{amsmath}
\usepackage{booktabs, caption, makecell}
\usepackage{threeparttable}

\newcommand{\atUCLA}{Dept. of Physics and Astronomy, Univ. of California, Los Angeles, Los Angeles, CA 90095.}
\newcommand{\atOSU}{Dept. of Physics, Center for Cosmology and AstroParticle Physics, Ohio State Univ., Columbus, OH 43210.}
\newcommand{\atUH}{Dept. of Physics and Astronomy, Univ. of Hawaii, Manoa, HI 96822.}
\newcommand{\atNTU}{Dept. of Physics, Grad. Inst. of Astrophys.,\& Leung Center for Cosmology and Particle Astrophysics, National Taiwan University, Taipei, Taiwan.}
\newcommand{\atKU}{Dept. of Physics and Astronomy, Univ. of Kansas, Lawrence, KS 66045.}
\newcommand{\atWashU}{Dept of Physics \& McDonnell Center for the Space Sciences, Washington Univ in St Louis, MO}
\newcommand{\atSLAC}{SLAC National Accelerator Laboratory, Menlo Park, CA, 94025.}
\newcommand{\atUD}{Dept. of Physics, Univ. of Delaware, Newark, DE 19716.}
\newcommand{\atUCL}{Dept. of Physics and Astronomy, University College London, London, United Kingdom.}

\newcommand{\atJPL}{Jet Propulsion Laboratory, Pasadena, CA 91109.}

\newcommand{\atChicago}{Dept. of Physics, Enrico Fermi Institute, Kavli Institute for Cosmological Physics, Univ. of Chicago , Chicago IL 60637.}

\newcommand{\atCalPoly}{Physics Dept., California Polytechnic State Univ., San Luis Obispo, CA 93407.}
\begin{document}

\title{Observation of an Unusual Upward-going Cosmic-ray-like Event in the Third Flight of ANITA}

\author{P.~W.~Gorham}
\affiliation{\atUH}

\author{B. Rotter}
\affiliation{\atUH}

\author{P.~Allison}
\affiliation{\atOSU}

\author{O.~Banerjee}
\affiliation{\atOSU}

\author{L.~Batten}
\affiliation{\atUCL}

\author{J.~J.~Beatty}
\affiliation{\atOSU}

\author{K. Bechtol}
\affiliation{\atChicago}

\author{K.~Belov}
\affiliation{\atJPL}

\author{D.~Z.~Besson}
\affiliation{\atKU}
\affiliation{National Research Nuclear Univ., Moscow Engineering Physics Inst., Moscow, Russia.}

\author{W.~R.~Binns}
\affiliation{\atWashU}

\author{V.~Bugaev}
\affiliation{\atWashU}

\author{P.~Cao}
\affiliation{\atUD}

\author{C.~C.~Chen}
\affiliation{\atNTU}

\author{C.~H.~Chen}
\affiliation{\atNTU}

\author{P.~Chen}
\affiliation{\atNTU}

\author{J.~M.~Clem}
\affiliation{\atUD}

\author{A.~Connolly}
\affiliation{\atOSU}

\author{L.~Cremonesi}
\affiliation{\atUCL}

\author{B.~Dailey}
\affiliation{\atOSU}

\author{C.~Deaconu}
\affiliation{\atChicago}

\author{P.~F.~Dowkontt}
\affiliation{\atWashU}

\author{B.~D.~Fox}
\affiliation{\atUH}

\author{J.~W.~H.~Gordon}
\affiliation{\atOSU}

\author{C.~Hast}
\affiliation{\atSLAC}

\author{B.~Hill}
\affiliation{\atUH}

\author{K.~Hughes}
\affiliation{\atOSU}
\altaffiliation{\atChicago}

\author{J.~J.~Huang}
\affiliation{\atNTU}

\author{R.~Hupe}
\affiliation{\atOSU}

\author{M.~H.~Israel}
\affiliation{\atWashU}

\author{A.~Javaid}
\affiliation{\atUD}

\author{J.~Lam}
\affiliation{\atUCLA}

\author{K.~M.~Liewer}
\affiliation{\atJPL}

\author{S.~Y.~Lin}
\affiliation{\atNTU}

\author{T.C. Liu}
\affiliation{\atNTU}

\author{A.~Ludwig}
\affiliation{\atChicago}

\author{L.~Macchiarulo}
\affiliation{\atUH}

\author{S.~Matsuno}
\affiliation{\atUH}

\author{C.~Miki}
\affiliation{\atUH}

\author{K.~Mulrey}  
\affiliation{\atUD}

\author{J.~Nam}
\affiliation{\atNTU}

\author{C.~J.~Naudet}
\affiliation{\atJPL}

\author{R.~J.~Nichol}
\affiliation{\atUCL}

\author{A.~Novikov}
\affiliation{\atKU}

\author{E.~Oberla}
\affiliation{\atChicago}

\author{M.~Olmedo}
\affiliation{\atUH}

\author{R.~Prechelt}
\affiliation{\atUH}

\author{S.~Prohira}
\affiliation{\atKU}

\author{B.~F.~Rauch}
\affiliation{\atWashU}

\author{J.~M.~Roberts}
\affiliation{\atUH}

\author{A.~Romero-Wolf}
\affiliation{\atJPL}

\author{J.~W.~Russell}
\affiliation{\atUH}

\author{D.~Saltzberg}
\affiliation{\atUCLA}

\author{D.~Seckel}
\affiliation{\atUD}

\author{H.~Schoorlemmer}
\affiliation{\atUH}

\author{J.~Shiao}
\affiliation{\atNTU}

\author{S.~Stafford}
\affiliation{\atOSU}

\author{J.~Stockham}
\affiliation{\atKU}

\author{M.~Stockham}
\affiliation{\atKU}

\author{B.~Strutt}
\affiliation{\atUCLA}

\author{G.~S.~Varner}
\affiliation{\atUH}

\author{A.~G.~Vieregg}
\affiliation{\atChicago}

\author{S.~H.~Wang}
\affiliation{\atNTU}

\author{S.~A.~Wissel}
\affiliation{\atCalPoly}

\vspace{2mm}
\noindent

\begin{abstract}
We report on an upward traveling, radio-detected cosmic-ray-like impulsive event with characteristics
closely matching an extensive air shower. This event, observed in the third
flight of the Antarctic Impulsive
Transient Antenna (ANITA), a NASA-sponsored long-duration balloon payload,
is consistent  with a similar event reported in a previous
flight. These events may be produced by the atmospheric decay of an upward-propagating $\tau$-lepton produced by a $\nu_{\tau}$ interaction,
although their relatively steep arrival angles create tension with the standard model (SM) neutrino cross section.
Each of the two events have {\it a posteriori} background estimates of $\lesssim 10^{-2}$ events. If these are generated
by $\tau$-lepton decay, then either the charged-current $\nu_{\tau}$ cross section is suppressed at EeV energies,
or the events arise at moments when the peak flux of a transient neutrino source was much larger than the typical
expected cosmogenic background neutrinos. 
\end{abstract}
\maketitle

The ANITA instrument is primarily designed for the detection of the ultra-high energy (UHE) cosmogenic neutrino flux via
the Askaryan effect in ice~\cite{Ask62,Sal01,T486}, but is able to trigger on a wide variety of different impulsive radio signals. 
During the first ANITA flight, an unanticipated radio signal was discovered: 16 events
due to ultra-high energy cosmic ray (UHECR) air showers were found during a blind search of the data for isolated non-anthropogenic events~\cite{ANITA_CR}.
ANITA observes UHECR via radio impulses that occur when geomagnetically-induced charged-particle 
acceleration occurs in the propagation of an extensive air shower in the atmosphere. Conventional
down-going ultra-high energy cosmic-ray (UHECR) air showers produce downward-propagating
radio impulses that are observed in reflection off the surface of the ice, leading to phase inversion 
of the signal. UHECR events detected by ANITA also include a subset of horizontally-propagating
stratospheric air showers seen just above the horizon, which point directly at the payload, and show no phase
inversion of the signal~\cite{Upshowers}. These observations have established a baseline for identification
of events of UHECR origin in ANITA data.

In the ANITA-I flight one such UHECR-like event was observed with
characteristics similar to the direct, horizontal cosmic rays, but from a direction well below the horizon, without the phase
inversion due to a reflection~\cite{Upshowers}. The background for this event was estimated to be $\leq 10^{-3}$ events, 
suggesting the possibility that such events could arise from a high-energy $\nu_{\tau}$
charged-current interaction in the ice, leading to a $\tau$-lepton which exits the ice surface and decays,
producing an air shower that propagates upward in the atmosphere. However, a possible anthropogenic
origin for the ANITA-I event could not be ruled out at sufficient confidence to be conclusive.

The third flight of the ANITA instrument took place
from Dec. 18, 2014 through Jan. 8, 2015, with 22 days at float at an altitude of $\sim 34$ to 38~km.
Unexpected strong continuous-wave (CW) interference from geosynchronous satellites limited
the effective full-payload exposure to about 7 days of equivalent time.
Despite this loss of sensitivity, a set of 20 radio-detected UHECR events were identified 
in a template-based analysis~\cite{Rotterthesis}.
Because the polarity of the events was the primary characteristic that would distinguish phase-inverted
events from the direct events, including possible upward-going showers, 
we blinded the event polarity throughout the analysis to avoid bias.
The geomagnetic field in Antarctic is predominantly vertical, and thus the
Lorentz-force acceleration of the $e^+e^-$ pairs in the shower leads to 
lateral charge-separation that produces an almost completely horizontally-polarized (Hpol) 
signal, with nearly unique temporal and spectral properties compared to anthropogenic
background events observed. Despite their small size, the residual horizontal 
components of the geomagnetic field still provide
for a detailed confirmation of the geomagnetic correlation of UHECRs. If we write the geomagnetic
field in a local cartesian basis, then $\mathbf{B} = (B_x, B_y, B_z)$, with
$B_x,~B_y \ll B_z$ as noted above.  ANITA's observation geometry also favors air showers with 
primary particle momenta with zenith angles of $60^{\circ}$ or more, and thus their longitudinal
velocity will follow $v_x,~v_y \gg v_z$ in general.

\begin{figure}[htb!]
 \includegraphics[width=3.2in]{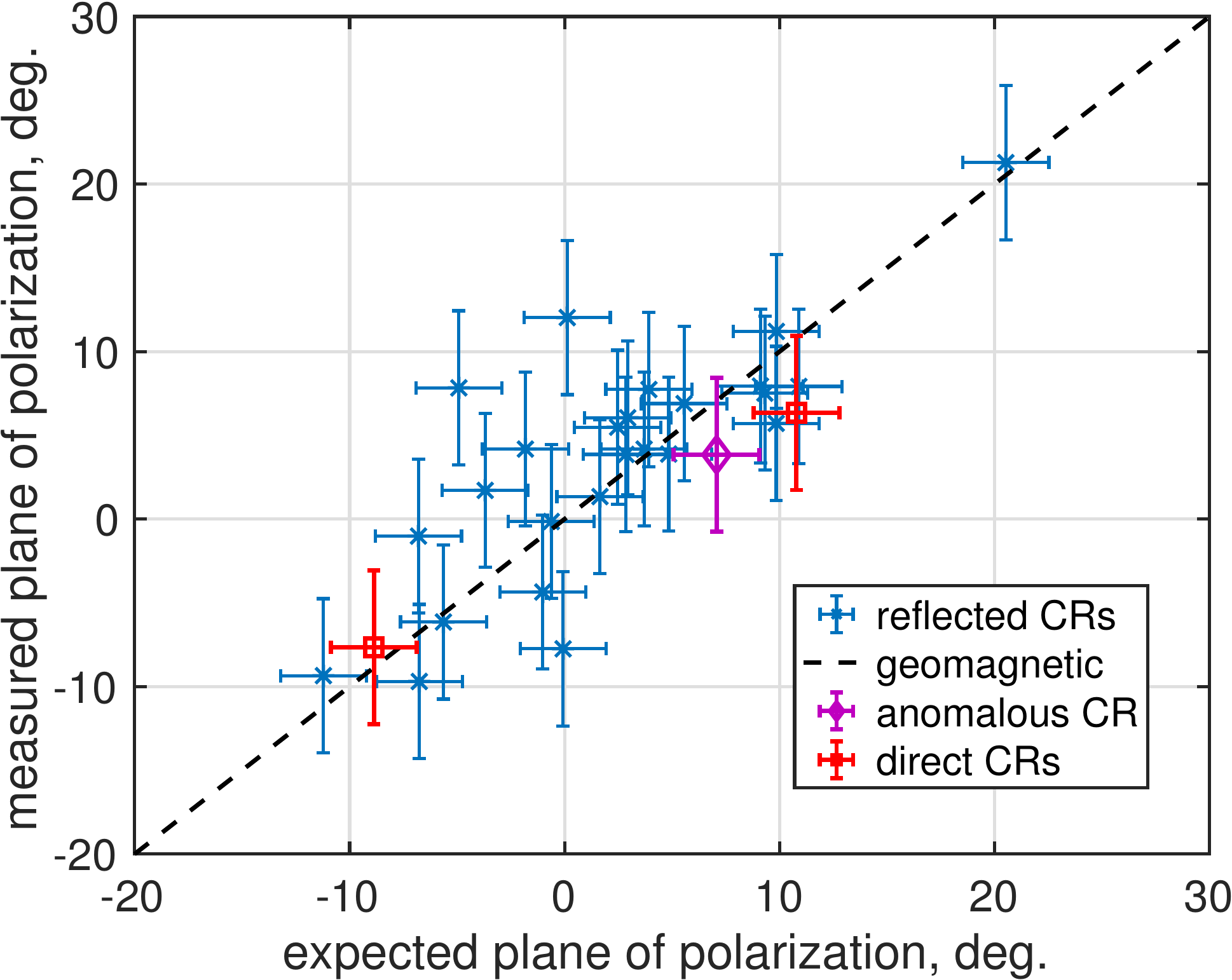}
 \caption{\it \small  Geomagnetic correlation of  20 UHECR events detected in ANITA-III, with
 event planes-of-polarization determined via Stokes parameters for each event. The 
 two above-horizon non-inverted CRs are shown in red, and the anomalous
 non-inverted, below-horizon CR-like event 15717147 is shown in magenta.
 \label{geomag}}
 \end{figure}

From Feynman's rule~\cite{Feynman}, 
the radiation field per particle will be aligned with the observer's apparent angular acceleration of the charge,
which is given by the magnetic portion of the Lorentz force,
$\mathbf{F} =  q~ \mathbf{v} \times \mathbf{B}.$
Neglecting terms that are second order in the acceleration, and recognizing that the magnetic deflection is 
nearly perpendicular to the direction of radiation, the observed radiation field vector can be approximated as
\begin{equation}
\mathbf{E} \propto (v_y B_z \hat{x} - v_xB_z \hat{y})  + (v_xB_y - v_yB_x) \hat{z}~.
\end{equation}
The first term in parentheses on the right hand side gives the 
Hpol component of the field, and because
it involves the strongest components of both $\mathbf{v}$ and $ \mathbf{B}$, it is the much 
stronger of the two radiation fields. The second term gives the vertically-polarized (Vpol) field component,
and is significantly weaker because it depends on the much weaker transverse magnetic field vector components.
In addition, there is a small contribution from Askaryan emission, but because of the strong Antarctic
geomagnetic field, this is limited to about 4\% of the total and is neglected here.
Because ANITA is designed to do accurate pulse-phase polarimetry with both 
Hpol and Vpol receiving antennas, the transverse $B$-field component is readily detectable. Since the geomagnetic
field is well-modeled in Antarctica, it provides a strong confirmation of geomagnetic association for a
given UHECR impulse, whereas signals of anthropogenic origin are uncorrelated to the geomagnetic field. 
Fig.~\ref{geomag} shows the geomagnetic-correlated results for the UHECR events selected in ANITA-III,
The expected polarization is corrected for the Fresnel coefficient of reflection where appropriate.
Measurement errors were determined by measurements of comparable calibration pulses, and 
include systematics.

\begin{figure}[htb!]
 \includegraphics[width=3.7in]{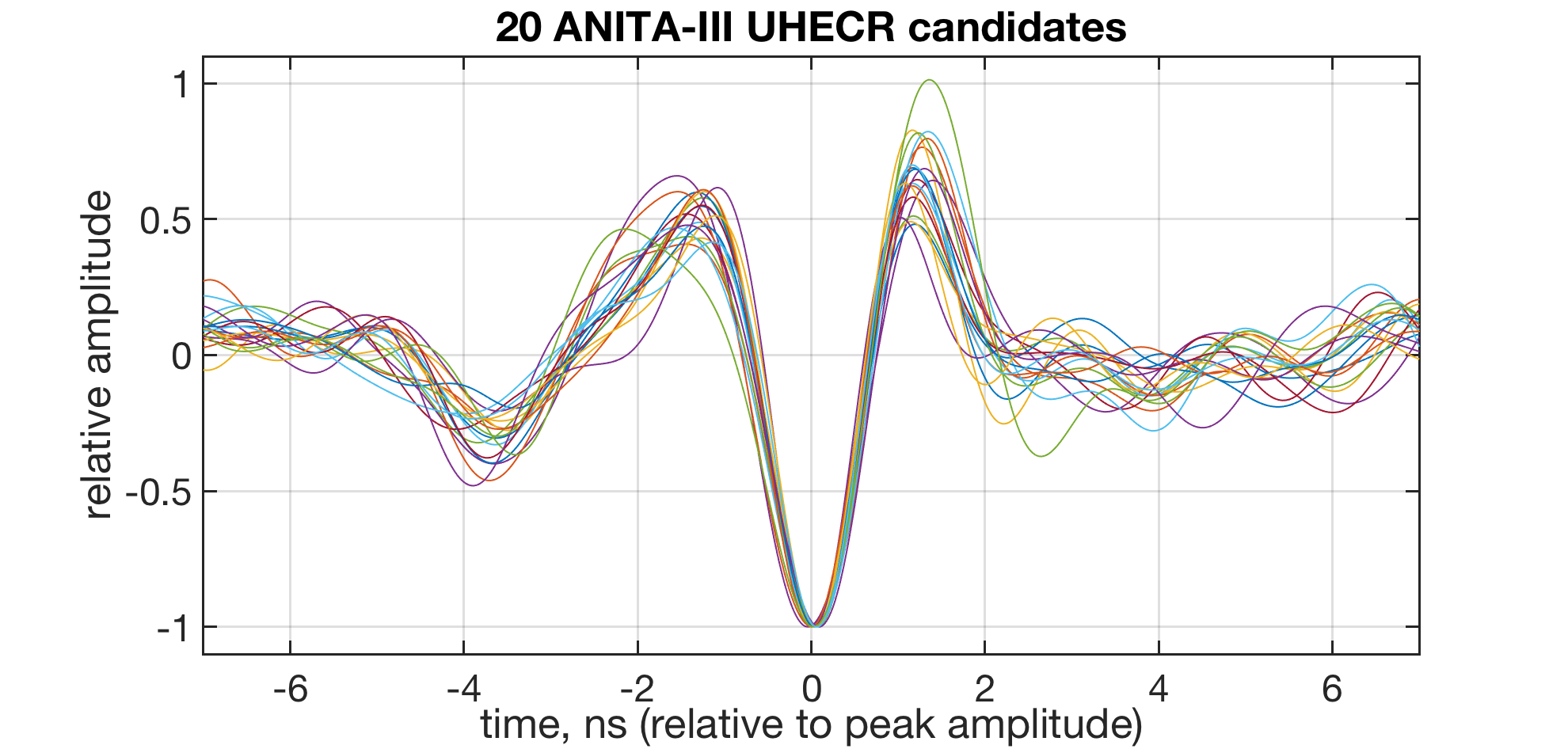}
 
 \caption{\it \small  Horizontally-polarized waveforms of 
 20 UHECR events detected in ANITA-III, with the polarity
 and amplitude all normalized to the peak.
 \label{overlay}}
 \end{figure}

The unblinded polarity of the ANITA-III CR events showed that the two above-horizon events 
among the sample had the expected non-inverted pulse phase,
consistent with their origin as stratospheric, atmosphere-skimming air showers. However, as noted above,
one of the remaining events also had a clearly non-inverted polarity, inconsistent with a reflection, but in all other
ways consistent with UHECR origin. Fig.~\ref{overlay} shows the overlain normalized Hpol
waveforms from each of the
20 candidate events, with the 17 inverted-polarity reflected events now un-inverted for direct comparison of the
waveform shape. The events have the instrumental response deconvolved, and are 
normalized in amplitude to their maximum magnitude. They are remarkably
similar in shape once the inversion is removed.

For the final 20-event UHECR selection, candidates were verified to be spatially and 
temporally isolated from any other events like them, and
showed a high degree of correlation with a waveform template determined by well-established models for
UHECR radio emission. We have identified no known physics backgrounds for these events. Potential background
comes from anthropogenic radio signals that might mimic the UHECR characteristics, or unknown processes
which might lead to non-inverted polarity on reflection from the ice; further investigation of polarity is
given in ref.~\cite{Suppl1}. Two independent background estimates for anthropogenic origin were made. 
The first, using the likelihood that the event was a statistical outlier of sub-threshold events within
its nearby locale, gave a background estimate of $B = 1.2 \times 10^{-3}$ events for the 20-UHECR sample~\cite{Rotterthesis}.  
The second method uses a probability for a single isolated UHECR-like background event, 
derived from the frequency of UHECR-like events that appeared in known anthropogenic
clusters of events and charted bases or camps. Because the rate of actual UHECR events is such
that some inevitably do get included (and therefore lost to the analysis) 
as part of these clusters, this latter estimate provides only an upper
limit to the background, $B \leq 0.015$ events, also for the entire 20 UHECR sample.
Thus by all indications the resulting selection of events represents a very pure sample of radio-detected UHECRs.

\begin{figure}[htb!]
 \includegraphics[width=3.2in]{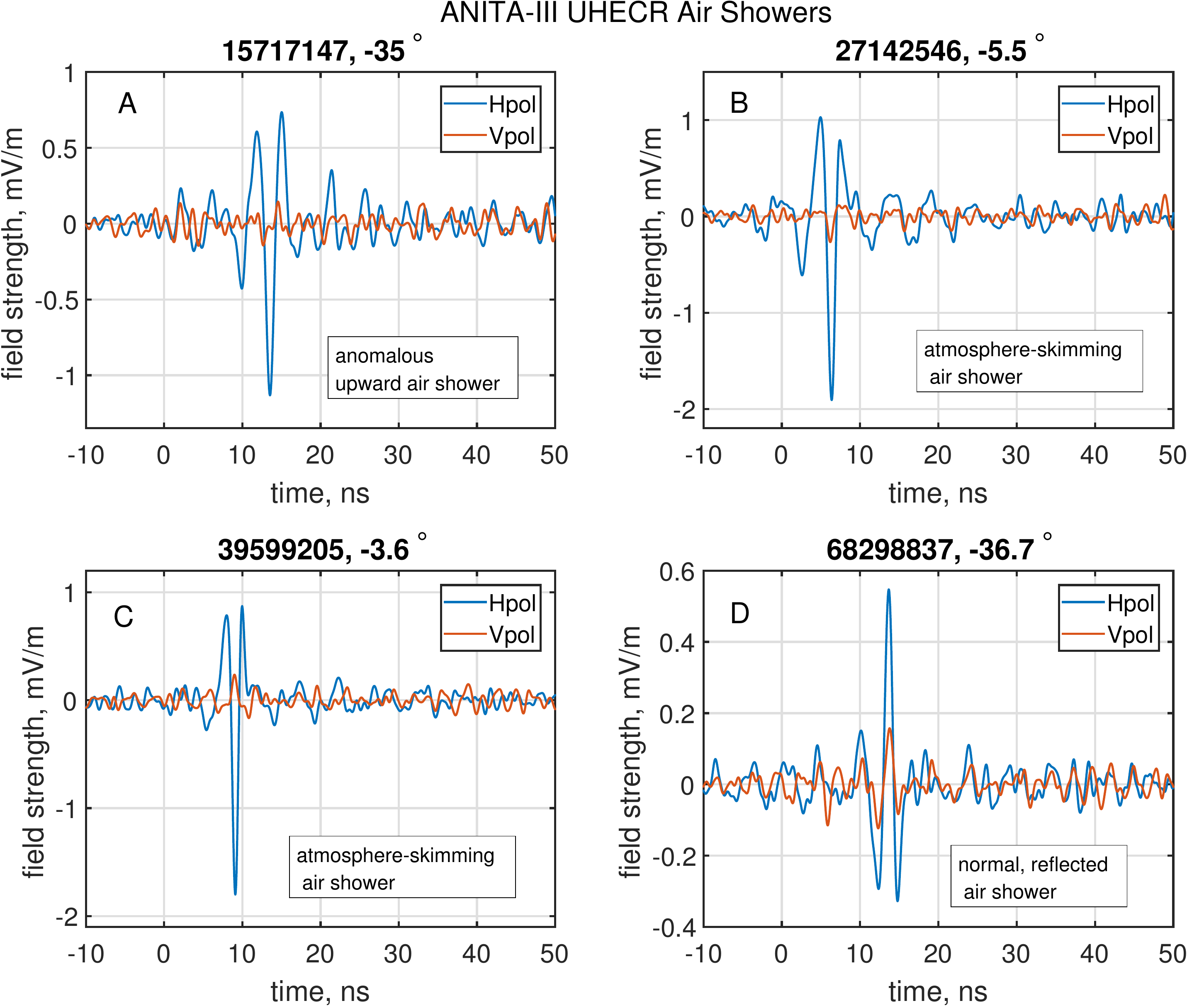}
 \caption{\it \small  The three non-inverted polarity events are shown in panels A,B,C. Panel A shows the anomalous event,
 with the same polarity as the above-horizon events B and C. Panel D shows the
 waveform for an inverted UHECR that had an upcoming angle close to that of the anomalous
 CR 15717147. The inversion of the normal reflected CR event is clearly evident.
 \label{CR4plot}}
 \end{figure}

Fig.~\ref{CR4plot} shows the incident field strength waveforms for all three of the events with non-inverted polarity, along
with one of the ``normal'' UHECR events, chosen because its arrival angle at the payload was similar to
that of the anomalous event 15717147.  
Detailed simulations of the UHECR radio emission
process find that the power spectral density (PSD) of the radio signal is dependent 
on the observer's viewing angle relative to the
axis of the air shower, and the PSD can thus be used, along with other parameters of the shower signal,
to  estimate the primary energy of the event~\cite{Harm16}. To provide more confidence in our
estimate, we cross-checked event 15717147 against 12 of the 16 ANITA-I cosmic ray events for which the
parameters could be directly compared and scaled. The results are quite consistent, yielding 
an estimated shower energy of $\mathit{E} = 0.56^{0.3}_{-0.2} \times 10^{18}$~eV for this event,
assuming that shower was initiated close to the event's projected position on the ice sheet. For
a shower initiated at a height of 4~km above the ice, the energy is reduced by about 30\% to
$\mathit{E} = 0.40$~EeV.  The errors here are statistical, based on the root-mean-square of the cross-check sample. 

In addition to the targeted search for UHECR events, we
performed two completely independent optimized multivariate blind analyses of all events, 
favoring impulsive, highly-linearly-polarized events, without consideration
of correlation to any UHECR waveform template~\cite{A3paper}. In both of these analyses, complete isolation from any
anthropogenic source or from any other events was a stringent requirement, and event 15717147 
passed in both cases. These two analyses confirm that
event 15717147 is unique, impulsive, and isolated, even when not selected by its UHECR-related properties. 
The {\it a posteriori} background estimates for both 15717147 and for the similar anomalous event
seen in ANITA-I~\cite{Upshowers} are at the $\gtrsim 3\sigma$ level. 
There is thus significant evidence for a physical process that leads to direct upward-moving cosmic-ray-like air 
showers above the ice surface.  

\begin{figure}[htb!]
 ~~~\includegraphics[width=3.25in]{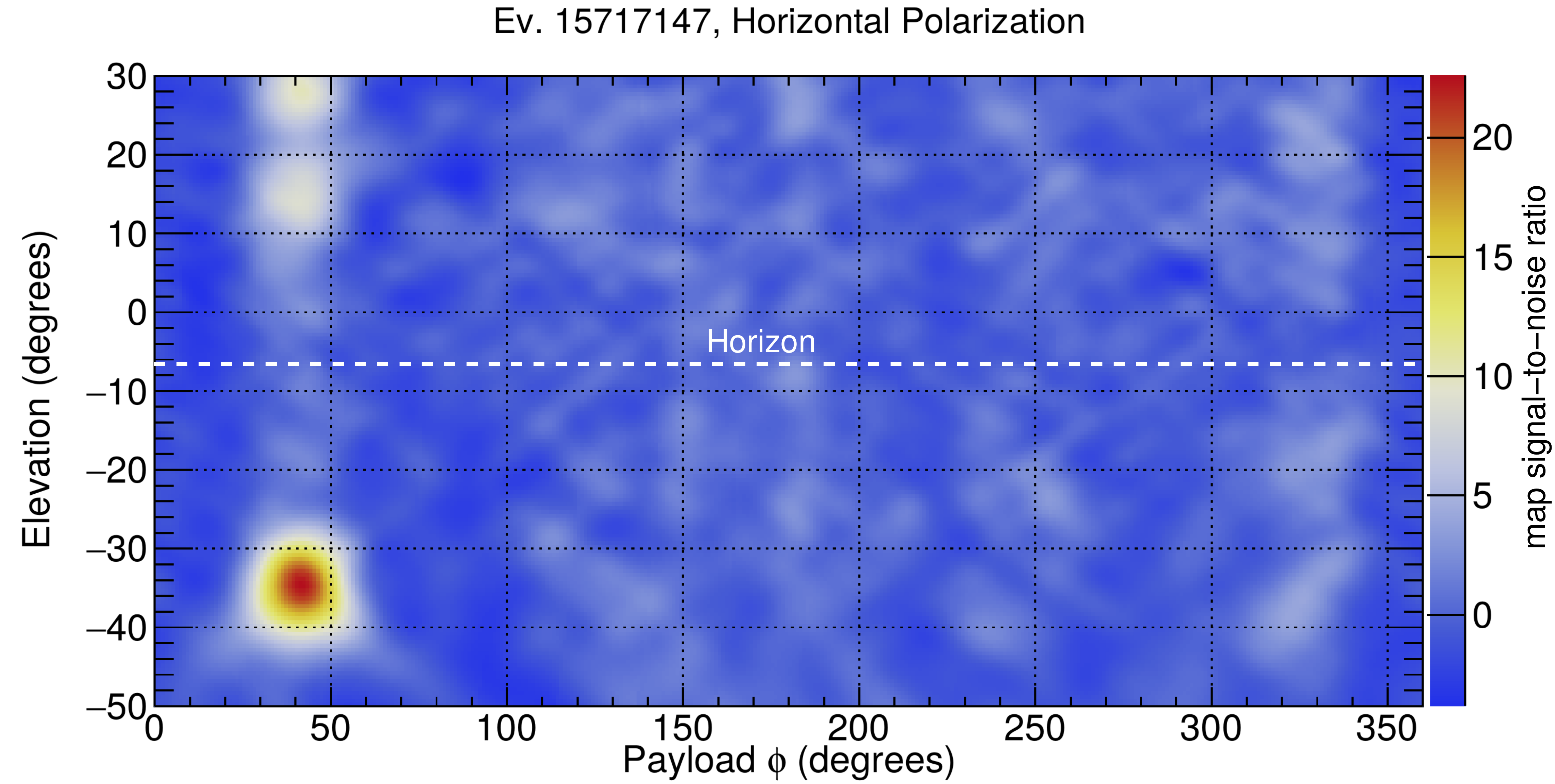}\\
  \includegraphics[width=3.0in,height=1.65in]{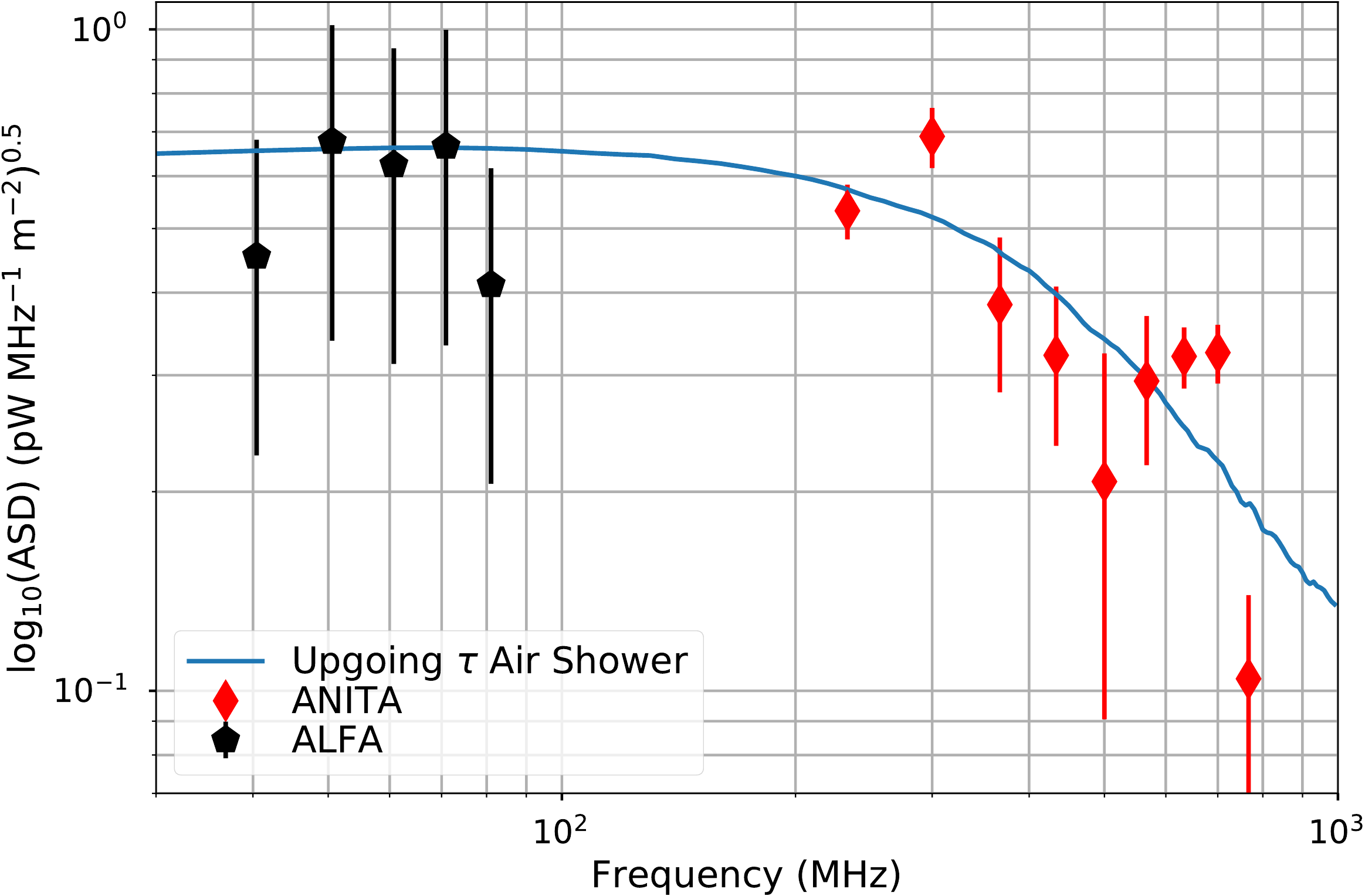}
 \caption{\it \small  Top: Interferometric map of the arrival direction of the anomalous CR event 15717147.
Bottom: ANITA combined amplitude spectral density (ASD) for the event, from 50-800 MHz, including data
from the ANITA Low Frequency Antenna (ALFA). A
simulated 
upward-propagating extensive air shower spectral-density curve is overlain.
  \label{map}}
 \end{figure}

For detected radio impulses, the large fields-of-view for the quad-ridged horns used in ANITA allow up to
15 antennas, drawn from up to 5 azimuthal sectors of the payload, to be used for coherent beam forming.
Pulse-phase interferometry between these antennas then yields a map of the arrival direction of the
radio impulse to typical precisions of $0.25^{\circ},~0.65^{\circ}$ in elevation and azimuth, respectively~\cite{ARW15}.
Fig.~\ref{map}(top) shows the resulting false-color map for event 15717147 in coordinates local to the payload,
scaled by the signal-to-noise ratio of the map. Elevation is with respect to the payload horizontal, and the azimuthal
angle $\phi$ is with respect to the payload heading at the event arrival time. Mapping is done 
for 360$^{\circ}$ in $\phi$ to verify that the beamforming solution is unique.


ANITA-III flew a separate low-frequency horizontally-polarized quad-slot antenna, the
ANITA low-frequency antenna (ALFA), covering the frequency
band from 30 to 80~MHz. ALFA's goal was to provide radio-spectral overlap of ANITA UHECR measurements with ground-based
data which generally favors bands below 100~MHz. Roughly 3/4 of the UHECR event sample reported here
were also detected in the ALFA, and of those detections, the ALFA data for 15717147 was among 
the events with the highest signal-to-noise ratio, in this case $\geq 5\sigma$ above the thermal noise. Fig.~\ref{map}(bottom)
shows the combined ASD for this event, including the ALFA data.
The overlain curve gives the simulated spectral density expected from a $\tau$-lepton initiated air shower,
with characteristics consistent with this event~\cite{Alv17}. While similar spectral density would be
expected for a normal CR air shower seen in reflection, these data which fit this non-inverted
event further strengthen its identification as an anomalous air shower.

An alternative explanation
of the similar ANITA-I event as due to transition radiation of an Earth-skimming event 
has also been proposed~\cite{Motloch17}. In this model, the plane-of-polarization
correlation to geomagnetic angles would be coincidental. Since the event observed in ANITA-III
is also well-correlated to the local geomagnetic angle, and both events are consistent within
3-5 degrees of measurement error, coincidental alignment for both appears
probable only at the few percent level. The waveform of these events showed a high degree
of correlation to radio-detected UHECRs in each flight, which supported their identification
as UHECRs.  Ref.~\cite{Motloch17} did not provide any detailed modeling 
of time-domain waveforms for transition radiation that confirm its
similarity to those made by the UHECR emission process. This step appears necessary before this hypothesis can
be further evaluated.

%
%

\begin{table}[htb!]
\begin{center}
\begin{footnotesize}
\begin{threeparttable}
\caption{\it \small ANITA-I,-III anomalous upward air showers.\label{tbl1}}
\begin{tabular}{|c|c|c|} \hline
 event, flight &  3985267, ANITA-I & 15717147, ANITA-III \\ \hline
 date, time  & 2006-12-28,00:33:20UTC  & 2014-12-20,08:33:22.5UTC \\
 Lat., Lon.$^{(1)}$ &  -82.6559, 17.2842 & -81.39856, 129.01626 \\
 Altitude  & 2.56~km & 2.75~km \\
 Ice depth & 3.53~km & 3.22~km \\
 El., Az. &$-27.4\pm0.3^{\circ},159.62\pm0.7^{\circ}$ & $-35.0\pm0.3^{\circ},61.41\pm0.7^{\circ}$\\
 RA, Dec$^{(2)}$& 282.14064, +20.33043  & 50.78203, +38.65498 \\
 $E_{shower}^{~~(3)}$ & $0.6 \pm 0.4~{\rm EeV}$ & $0.56^{+0.3}_{-0.2}~{\rm EeV}$\\  \hline
\end{tabular}
\begin{tablenotes}
\item[1] Latitude, Longitude of the estimated ground position of the event.
\item[2] Sky coordinates projected from event arrival angles at ANITA.
\item[3] For upward shower initiation at or near ice surface.
\end{tablenotes}
\end{threeparttable}
\end{footnotesize}
\end{center}
\end{table}%

Table~\ref{tbl1} gives measured and estimated parameters for both of the anomalous CR events,
with sky coordinates derived from the arrival direction of the radio impulses.

In our report of the ANITA-I anomalous CR event, we considered the hypothesis that such events could
arise through decay of emerging $\tau$-leptons generated by $\nu_{\tau}$ interactions beneath the ice surface.
However, the interpretation of these events as $\tau$-lepton decay-driven air showers, arising from a diffuse flux of
cosmic $\nu_{\tau}$, faces the difficult challenge that the chord lengths through the Earth are such that
the Standard Model (SM) neutrino cross section~\cite{CTS11}, even including the effect of $\nu_{\tau}$
regeneration~\cite{regen}, will attenuate the flux by a factor of $10^{-5}$~\cite{Andres17,Alv17}. 
Event 15717147 emerged from the ice with a zenith angle of $\sim 55.5^{\circ}$, implying a chord
distance through the Earth of $\sim 7000$~km, or $3 \times 10^4$ km water equivalent, 
a total of 18 SM interaction lengths at 1~EeV. Even with combined effects of $\nu_{\tau}$
regeneration, and significant suppression of the 
SM neutrino cross section above $\sim 10^{18}$~eV, an alternative model, such as a strong
transient flux from a source with compact angular extent, is required to avoid exceeding current bounds
on diffuse, isotropic neutrino fluxes.

Suppression of the cross section may occur even within the SM for the extremely low values of the
Bjorken-{$x$} parameter that obtain at ultra-high energies. For example, ref.~\cite{Henley06}
shows examples where higher-than-expected gluon saturation at $x  < 10^{-6}$ causes the
UHE deep-inelastic neutrino cross section to saturate at $10^{18}$~eV,  remaining essentially constant above
that energy. This yields a factor of 3-4 suppression compared to the SM at $10^{19}$ eV,
approaching an order of magnitude at $10^{20}$~eV. More recent studies show
similar types of suppression are possible, giving factors of 2-3 at $10^{18-19}$~eV~\cite{Armesto08, Ill2011}.
Such SM-motivated scenarios would certainly decrease the exponential attenuation for
the Earth-crossing neutrinos relevant to our case, but unless the suppression is an order of magnitude or
more, a large transient point-source flux is likely still required. Thus we consider also
a search for potential candidate transients that may be associated with this event.


Under the hypothesis that event 15717147 is a $\tau$-lepton-initiated air shower, 
the angular error relative to the
parent neutrino direction is $\sim 1.5^{\circ}$, arising from both the 
width of the emission cone~\cite{Harm16}, and
the instrinsic statistical errors in our estimate of the arrival direction of the RF signal. 
To investigate this hypothesis further, we point back along the apparent arrival direction, giving
sky coordinates shown in Table~\ref{tbl1}. With these parameters, we search existing catalogs 
for associations with two transient source types for which source confusion 
is not excessive: gamma-ray burst (GRB) sources, and supernovae.
GRBs have been considered as possible UHE neutrino sources for many years,
although there are no detections to date. Supernovae (SNe) have also been proposed as
UHE sources in a variety of scenarios, both in core-collapse SNe, and more recently
even in type Ia SNe, which are believed to originate in the ignition of a white dwarf (WD)
progenitor. In the latter case, tidal ignition of a WD by interaction with an intermediate-mass
black hole has been proposed as a potential source of UHECRs~\cite{UHESN1,UHESN2,UHESN3}.

For the 1.5$^{\circ}$ radius error circle derived from the angular emission pattern
for UHECR events, no concurrent GRBs are observed.
A SN candidate is found to be associated: 
SN2014dz, a nearby type Ia SN at $z = 0.017$, is within $1.19^{\circ}$,
well within our expected angular uncertainty on the sky.
This relatively bright SN was discovered $\sim 7$ days before maximum, on 2014-12-20.146~\cite{2014dz}.
Our event time follows the initial discovery by just over five hours. 
Using  catalogued SNe discoveries during our flight, and a Bayesian estimator~\cite{Suppl1}, we find
the {\it a posteriori} probability of a chance association with any confirmed SN, at any redshift, within the estimated
likely time period of detectability for this SN, is $P \simeq 3.4 \times 10^{-3}$, or $2.7\sigma$. 

If SN2014dz is the source of the putative neutrino candidate, the implied peak isotropic neutrino luminosity must likely
far exceed the estimated bolometric luminosity of 
$L_B = 4.4 \times 10^{42}$~ergs~s$^{-1}$. The lower limit comes already from assuming a much lower
cross section than the SM. Alternatively, a beaming hypothesis would significantly relax these constraints.

Both the IceCube~\cite{ICtau} and Auger observatories are sensitive to $\tau$-leptons, IceCube through events transiting the
detector, or via $\tau-$decay within the detector, and Auger via Earth-skimming $\tau-$decay-initiated air
showers within a few degrees of the horizon~\cite{Auger15}. In this case, the declination for IceCube implies an additional
$\sim 4300$~km water equivalent column density, but if the SM cross section is suppressed, the $\sim 1$~km$^2$
geometric area of IceCube is still comparable to ANITA's effective point-source geometric area of $\sim 4$~km$^2$ at
this arrival angle. Auger has potentially a much larger effective point-source area, but only limited exposure around
the time of our event. 
However if the transient flux was as large as it appears, coincident detections in archival data
may be possible.

A search of the projected position given by the similar anomalous event from ANITA-I in 2006 yielded 
no SNe or any other significant association,  but the sky position for this event 
is within $\sim 10^{\circ}$ from the galactic plane, and thus extinction
leads to low SNe detection efficiency for this region of the sky.

We thank NASA for their generous support of ANITA, and the Columbia
Scientific Balloon Facility for their excellent field support, and the National Science
Foundation for their Antarctic operations support. This work
was also supported in part by the US Dept. of Energy, High Energy Physics
Division.

\end{document}